\documentclass{article}
\usepackage{graphicx}
\usepackage{subcaption}

\usepackage{layout}
\usepackage{multicol}
\usepackage{float}

\usepackage[font=it]{caption}
\title{\vspace{-2cm}Realtime Spectrum Monitoring via Reinforcement Learning --\\ A Comparison Between Q-Learning and Heuristic Methods}
\author{Tobias Braun$^1$, Tobias Korzyzkowske$^1$, Larissa Putzar$^1$, Jan Mietzner$^1$, and Peter A. Hoeher$^2$\\$^1$Hamburg University of Applied Sciences$\;\;$ $^2$Kiel University\\[1ex] E-mail (corresponding author): tobias.braun@haw-hamburg.de}
\date{July 11, 2023\vspace{-0.5cm}}

\makeatletter

\renewcommand\section{\@startsection {section}{1}{\z@}
                                   {0.1cm}
                                   {0.1cm}
                                   {\normalfont\Large\bfseries}}
\renewcommand\subsection{\@startsection {subsection}{1}{\z@}
                                   {0.1cm}
                                   {0.1cm}
                                   {\normalfont\large\bfseries}}
\makeatother
    
\begin{document}

	\maketitle
	
	\begin{multicols}{2}
	
	\section*{Abstract}
	
	\noindent Due to technological advances in the field of radio technology and its availability, the number of interference signals in the radio spectrum is continuously increasing. Interference signals must be detected in a timely fashion, in order to maintain standards and keep emergency frequencies open. To this end, specialized (multi-channel) receivers are used for spectrum monitoring. In this paper, the performances of two different approaches for controlling the available receiver resources are compared. The methods used for resource management (ReMa) are linear frequency tuning as a heuristic approach and a Q-learning algorithm from the field of reinforcement learning. To test the methods to be investigated, a simplified scenario was designed with two receiver channels monitoring ten non-overlapping frequency bands with non-uniform signal activity. For this setting, it is shown that the Q-learning algorithm used has a significantly higher detection rate than the heuristic approach at the expense of a smaller exploration rate. In particular, the Q-learning approach can be parameterized to allow for a suitable trade-off between detection and exploration rate.
	
	\section{Introduction}
	
	\noindent The use of radio signals in areas such as smart factories, smart homes, or autonomous driving is constantly increasing the number of radio signal transmitters. Improper handling leads to an increase in unwanted radio emissions. The usable spectrum for mobile radio applications is being expanded by new technologies such as 5G. At the same time, unwanted signals are more difficult to detect when the overall bandwidth of the frequency spectrum to be monitored keeps growing. This increases the vulnerability of wireless networks and services due to unwanted interference signals. Therefore, in order to enforce radio standards and to keep emergency frequencies open, the frequency spectrum must still be monitored in a timely fashion. In particular, an active search for interference signals is required so that they can be detected, classified, and localized early (interference hunting). 
	
	Frequency spectrum monitoring (FSM) systems typically consist of several receiver channels and corresponding software to control these channels. Specifically, the instantaneous bandwidths of the receiver channels cover only a small part of the overall spectrum to be monitored (due to cost reasons). Therefore, the aim is efficiently monitor the entire spectrum of interest with only few receiver channels by shifting the frequency windows of the receiver channels according to a suitable switching pattern. By this means, the entire spectrum can be visited over subsequent measurements.
	Controlling the frequency windows of the individual receiver channels is a typical problem of resource management (ReMa): Few receiver channels have to cover a relatively broad frequency spectrum within a limited time frame. By varying the frequency windows, the detection of continuous interference signals is relatively easy, while the probability of detecting pulsed signals is relatively low. State-of-the-art FSM systems use heuristic approaches to shift the frequency windows, which do not necessarily derive from any formal optimization procedure.
	
	\subsection{State of the Art}
	
	\noindent In the area of spectrum monitoring, previous heuristic approaches for ReMa can react in an agile fashion to current events within a certain framework. ReMa can be centralized [1] on a single network controller or decentralized [2] employing several devices. In other areas of communications technology involving similar ReMa problems, machine learning methods have already been successfully applied. For example, reinforcement learning methods can be used to optimize ReMa in mobile communication systems [3], taking into account aspects such as guaranteed data rates and fairness among users. However, such problem setups -- or related setups regarding optimal user scheduling -- are quite different from ReMa problems encountered in spectrum monitoring, so that corresponding solutions cannot simply be adopted or tailored to the specific FSM system contexts and associated requirements. Some related works, e.g. [4], indeed combine spectrum monitoring and machine learning -- specifically convolutional neural networks~-- but these focus on signal classification tasks rather than on ReMa, which involves a fundamentally different problem setup.
	
	\subsection{Paper Outline}
	\noindent In the following, we investigate the performances of two different ReMa approaches for spectrum monitoring, namely linear frequency tuning as a heuristic approach and a Q-learning algorithm from the field of reinforcement learning. To test the methods to be investigated, a simplified scenario was designed with two receiver channels monitoring ten non-overlapping frequency bands with non-uniform signal activity. For this setting, we demonstrate that the Q-learning algorithm used has a significantly higher detection rate than the heuristic approach at the expense of a smaller exploration rate. As will be seen, a particular advantage of the Q-learning approach is that it can be parameterized to allow for a suitable trade-off between detection and exploration. 
	
	The remainder of the paper is organized as follows: The environment for the Q-learning algorithm is defined in Section 2. The generation of training data is described in Section 3, and the two ReMa approaches are in detail described in Section 4. Numerical performance results are analyzed in Section~5, and conclusions are drawn in Section 6.
	
	\section{Environment}
	
	\noindent In order to calculate detection rates of the methods under investigation in the analysis part, the actual number of all interference signals within an observation series (OS) must be known. To this end, we conducted a computer simulation study for a simplified scenario with two receiver channels and employed generated test data. In the considered scenario, an OS consisted of 100 observed frequency spectra. The overall frequency spectrum was divided into 10 non-overlapping sub-spectra. For the sub-spectra, normalized frequency values were employed, as the exact size, start, and end of the associated frequency range are irrelevant for the ReMa problem. The number of generated interference signals was chosen as three throughout our simulation study. 
	
	The interference signals were continuous signals and had a 50 percent probability of being found in the first three sub-spectra. In particular, multiple interference signals could be located within the same sub-spectrum. In an observed spectrum of an OS, an interference signal had an 80 percent chance of being detected by one of the two receiver channels. This means that in 20 percent of the cases, an interference signal could not be detected by a receiver. The receivers were able to change their frequency range between the sub-spectra in an instantaneous fashion, i.e., the change of the frequency range did not entail any delay. Furthermore, the time required to observe and detect an interference signal over the entire spectrum was assumed to be equal to one time step (i.e., a single observed frequency spectrum) in the OS. 
	
	The feedback of the detection is a binary signal: a successful detection of an interference signal is represented by a one, an unsuccessful detection by a zero. Thus, the model does not have any information about the probability of whether an interference signal has been detected or not, only the binary classification into detected and non-detected interference signals is available. A false-positive detection of an interference signal was not considered within the scope of this paper for simplicity, as it has little relevance for the considered ReMa problem. 
	
	In our simulation setup, the signal occurrence probability is a free variable and can be varied over several measurements. Within one OS, the detection probability and the signal occurrence probability were kept constant for all observed frequency spectra. The number of generated interference signals, the number of sub-spectra within the overall frequency spectrum, and the number of receiver channels were also kept constant. As such, these values represent a simplified spectrum monitoring scenario with central control of the receiver channels and are based on  empirical experience derived from current FSM systems.
	
	\begin{figure*}[t]
	    \vspace{-0.5cm}
	    \begin{minipage}{0.475\linewidth}
	        \hspace{0.4cm}
	        \includegraphics[width=0.85\linewidth]{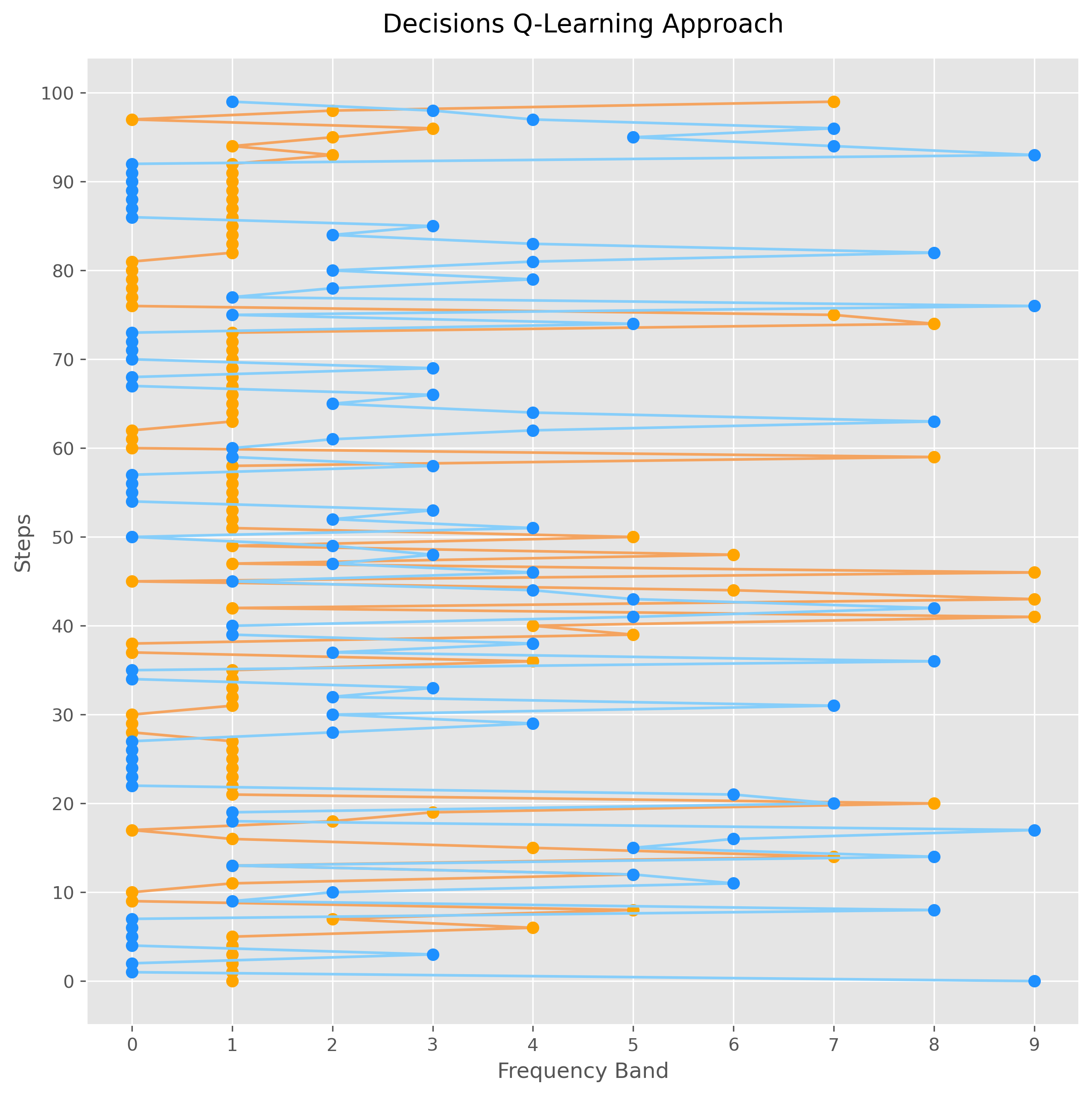}
	        \vspace{-0.3cm}
	        \captionof{figure}{Positions of receiver channels, set via the Q-learning algorithm with 0.2 exploration rate.\\[3ex]}
	        \label{fig:Q_Learning_20}
	    \end{minipage}
	    \hspace{0.75cm}
	    \begin{minipage}{0.475\linewidth}
	        \hspace{0.4cm}
	        \includegraphics[width=0.85\linewidth]{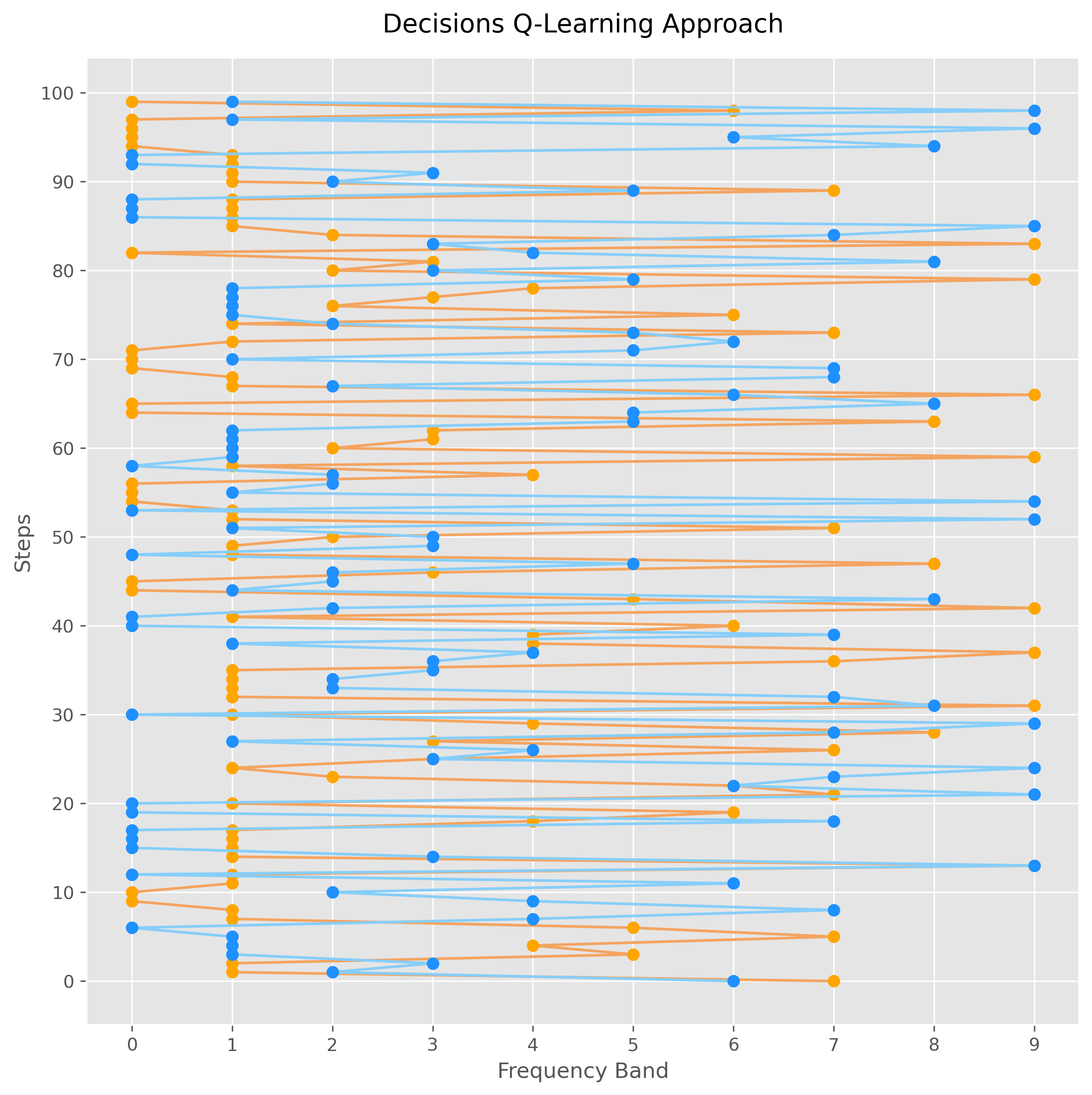}
	        \vspace{-0.3cm}
	        \captionof{figure}{Positions of receiver channels, set via the Q-learning algorithm with 0.5 exploration rate.}
	        \label{fig:Q_Learning_50}
	    \end{minipage}
	    \newline
	    \begin{minipage}{0.475\linewidth}
	        \hspace{0.4cm}
	        \includegraphics[width=0.85\linewidth]{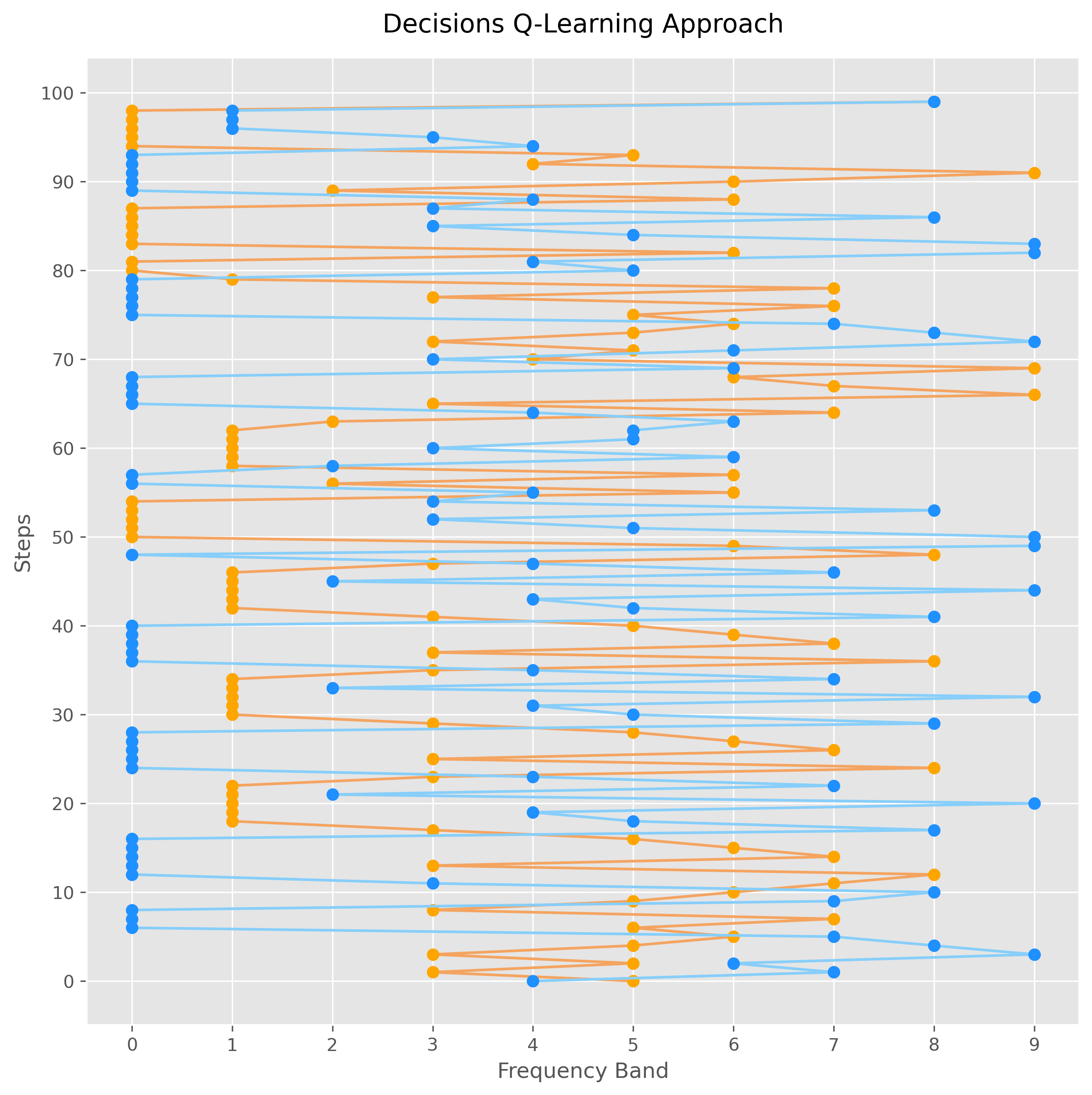}
	        \vspace{-0.3cm}
	        \captionof{figure}{Positions of receiver channels, set via the Q-learning algorithm with memory.\\[3ex]}
	        \label{fig:Q_Learning_Memory}
	    \end{minipage}
	    \hspace{0.75cm}
	    \begin{minipage}{0.475\linewidth}
	        \hspace{0.4cm}
	        \includegraphics[width=0.85\linewidth]{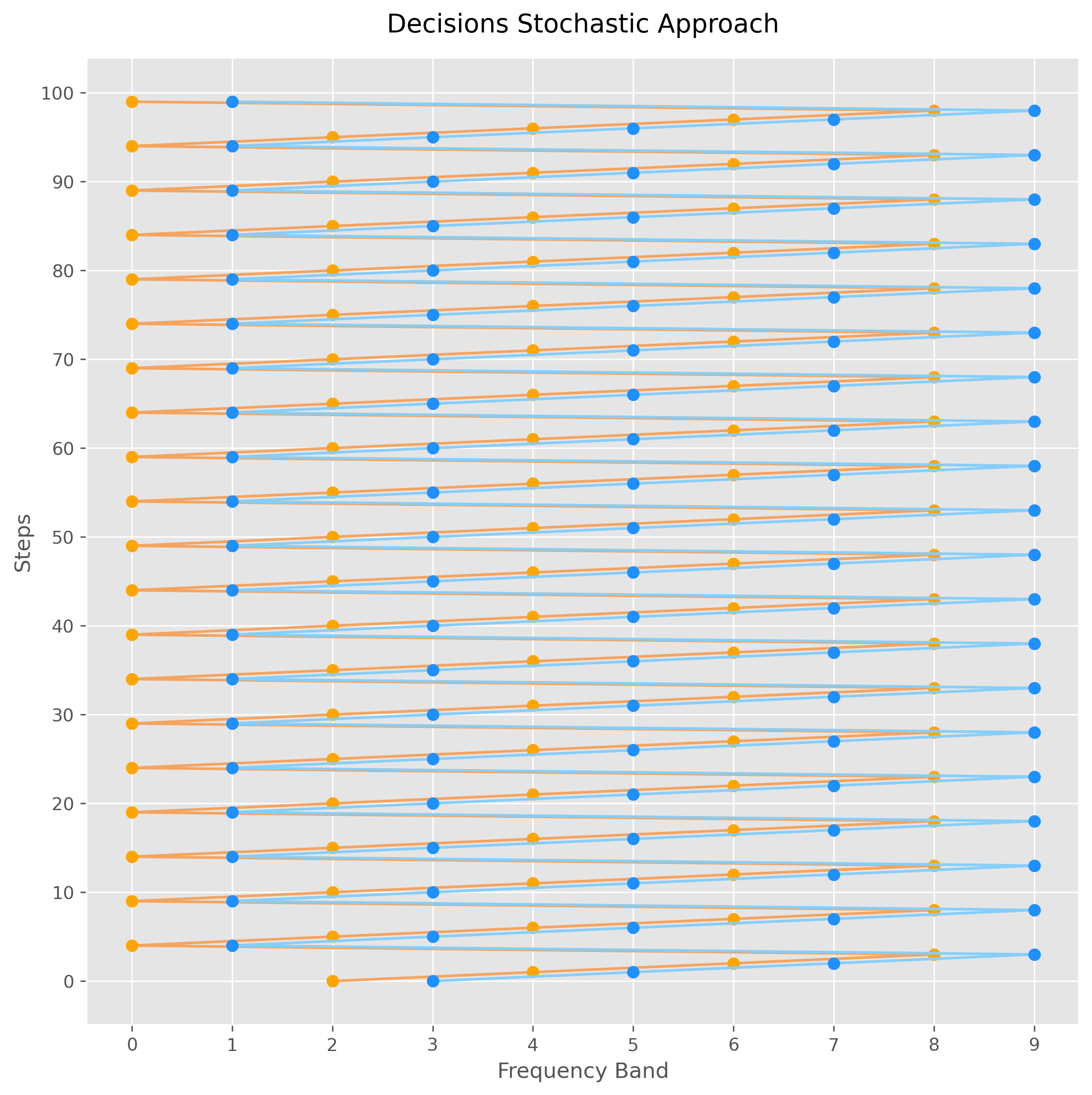}
	        \vspace{-0.3cm}
	        \captionof{figure}{Positions of receiver channels, set via linear frequency tuning.}
	        \label{fig:Lin_Tuning}
	    \end{minipage}
	    \vspace{-0.1cm}
	\end{figure*}
	
	\section{Generating Data}
	
	\noindent In accordance with the environment, the training and validation data were generated with 100 spectra per OS and 10 sub-spectra per observed frequency spectrum. This results in a list of 100 x 10 elements composed of zeros and ones for the binary feedback of the signal detection step. Since the data are generated, any amount of test data could be produced. Therefore, a 50/50 split between training and validation data, with 10,000 observation series each, was used.
	
	\section{Algorithms}
	
	\noindent To analyze the two approaches, the algorithms were implemented in Python. For the training of the Q-learning algorithm, 10,000 OS were used. The heuristic linear frequency tuning follows a fixed frequency-switching scheme and thus does not rely on training data.
	
	\subsection{Reinforcement Learning}
	
	\noindent The Q-learning algorithm is a model of the Markov decision process. For each possible state of the system, values (Q-values) for the prospects of success of the possible actions are stored in a Q-table. The states are stored as the status of the system and, together with the number of possible actions, determine the dimensionality of the Q-table. According to the environment, the agent (Q-agent) has the option to set each receiver channel to 10 sub-spectra for each state of the model. This results in 10 possible actions per receiver channel.
	A state of the model includes the following information:
	
	\vspace{-0.1cm}
	\begin{itemize}
	    \setlength\itemsep{-0.4em}
	    \raggedright
	    \item Current positions of the two receiver channels
	    \item Outputs of the interference signal detection from the previous spectrum within the OS.
	\end{itemize}
	\vspace{-0.1cm}
	
	%\noindent Die Q-Tabelle wird mit zufälligen Q-Values initiali-siert. Die Q-Values werden von dem Q-Agent ausgetestet und anhand von festgelegten Regeln aktualisiert:
	
	\noindent The Q-table is initialized with random Q-values. The Q-values are tested by the Q-agent and updated according to predetermined rules:
	
	\vspace{-0.1cm}
	\begin{itemize}
	    \setlength\itemsep{-0.4em}
	    \raggedright
	    %\item Verschlechterung des Q-Values, wenn beide Empfänger auf die gleiche Position gesetzt wurden.
	    %\item Verschlechterung des Q-Values, wenn beide Empfänger die Positionen getauscht haben.
	    %\item Verschlechterung des Q-Values, wenn kein Störsignal detektiert wurde.
	    %\item Verbesserung des Q-Values, wenn ein Störsignal detektiert wurde. Der Bonus steigt mit wiederholter Erkennung bis zu 5 mal.
	    \item Degradation of the Q-value, if both receiver channels were set to the same position.
	    \item Degradation of the Q-value, if both receiver channels have swapped positions.
	    \item Degradation of the Q-value, if no interference signal is detected.
	    \item Improvement of the Q-value, if an interference signal is detected. 
	    \item This bonus increases with repeated detection up to $x$ times, where we chose $x=5$ throughout.
	\end{itemize}
	\vspace{-0.2cm}
	
	\begin{figure*}[t]
	    \vspace{-0.5cm}
	    \begin{minipage}[b]{0.475\linewidth}
	        %\hspace{0.9cm}
	        \includegraphics[width=0.99\linewidth]{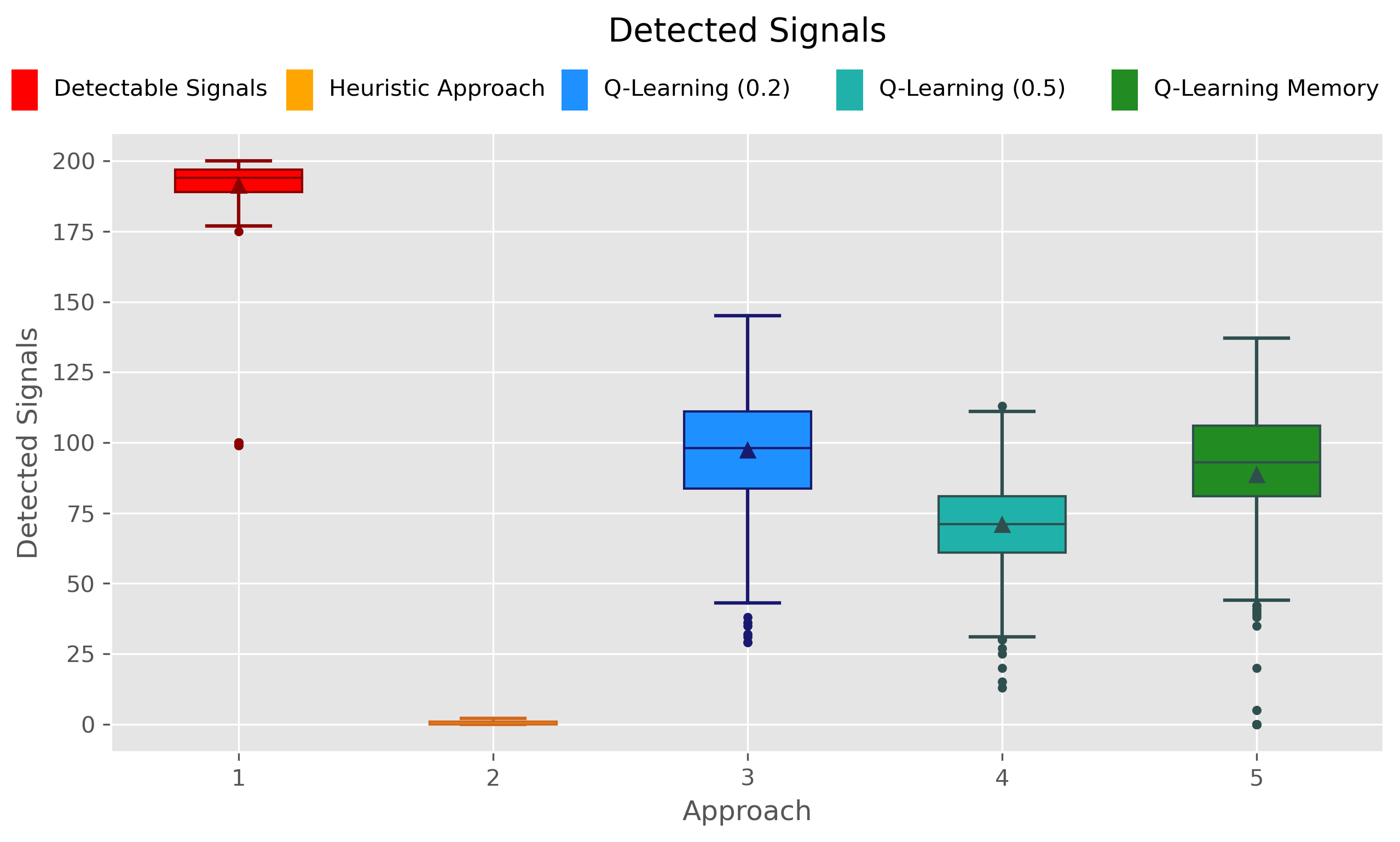}
	        \vspace{-0.3cm}
	        \captionof{figure}{Bar graph displaying detected signals.}
	        \label{fig:Analyses_Detections}
	    \end{minipage}
	    \hspace{0.75cm}
	    \begin{minipage}[b]{0.475\linewidth}
	        %\hspace{0.9cm}
	        % 2 Versions: with/without flyers
	        \includegraphics[width=0.99\linewidth]{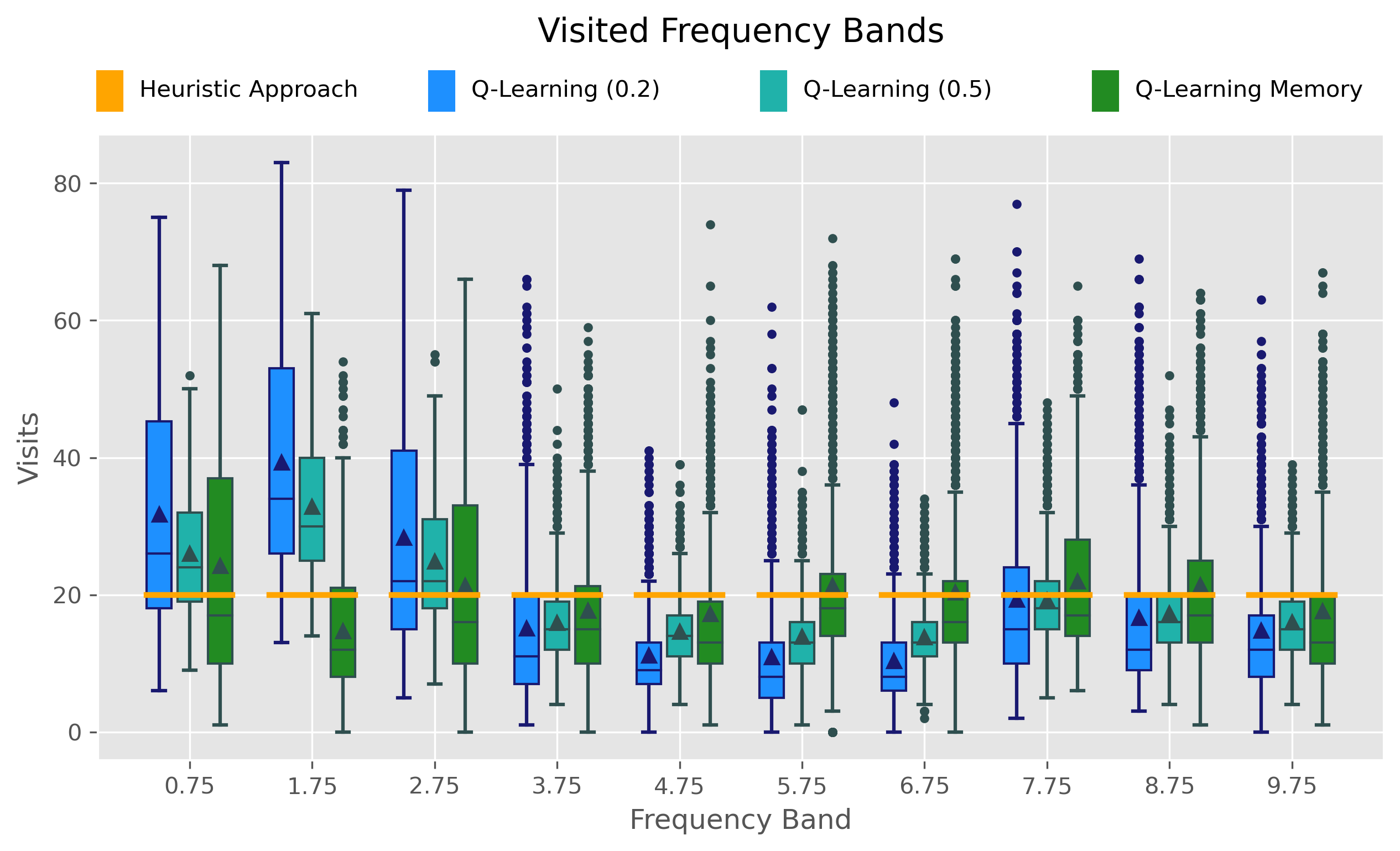}
	        \vspace{-0.3cm}
	        \captionof{figure}{Bar graph displaying visited frequency bands.}
	        \label{fig:Analyses_Visits}
	    \end{minipage}
	    \vspace{-0.2cm}
	\end{figure*}
	
	\noindent The Q-learning algorithm learns, based on the current positions of the receiver channels and the interference signal detection of the last step, how promising different distributions of the receiver channels are for the next step. If an interference signal has been detected, a receiver channel observes the interference signal, until it is terminated. Meanwhile, the second receiver channel continues to scan the spectrum. The Q-agent can transfer structures of the training data set into the Q-table. This way, it is learned, on which partial spectra interference signals occur more frequently. An exploration value can be defined separately. This can force the Q-agent to terminate the repeated interference signal observation and to continue exploration of other partial spectra. Examples for the behavior of the Q-learning algorithm regarding the positioning of the receiver channels under different exploration values are presented in Fig. \ref{fig:Q_Learning_20} and Fig. \ref{fig:Q_Learning_50}.
	
	For better control of when the Q-agent should switch the sub-spectrum after a detection series, a variant of the Q-learning algorithm with memory was developed. Along with the positions and detection properties of the receiver channels, a number of memory steps for both receiver channels are stored in the status. From this, a new rule for updating the Q-values can be created:
	
	\vspace{-0.1cm}
	\begin{itemize}
	    \setlength\itemsep{-0.4em}
	    \raggedright
	    \item Deterioration  of the Q-value, when an interference signal has been detected on the same sub-spectrum more than $x=5$ times in a row.
	\end{itemize}
	\vspace{-0.1cm}
	
	\noindent This rule prompts the Q-agent to terminate the observation after $x=5$ consecutive observations of an interference signal on the same sub-spectrum. This allows the Q-agent to learn to continue with the exploration after $x=5$ detections. An exemplary behavior of the Q-learning algorithm with memory regarding the positioning of the receiver channels is shown in Fig. \ref{fig:Q_Learning_Memory}.
	
	\subsection{Heuristic Approach}
	
	\noindent In the heuristic approach, the two receiver channels are placed side by side at the first and second sub-spectrum for each new OS. In each time step (observed frequency spectrum), the receiver channels are moved up by two sub-spectra. This repeats until the right end of the overall frequency spectrum is reached. Once the right end is attained, the receiver channels are reset to their initial positions at the first and second sub-spectrum. Since this approach is deterministic, it is not trained with any training data. An exemplary behavior of the heuristic algorithm regarding the positioning of the receiver channels is shown in Fig. \ref{fig:Lin_Tuning} as a reference.
	
	\section{Analysis}
	
	\noindent Since more interference signals can be generated during the data creation than can be detected with two receiver channels, a 100-percent detection rate (DR) is not possible. The detection rate of the approaches can only be compared with the number of detectable interference signals, shown in Fig.~\ref{fig:Analyses_Detections} in red color. The heuristic approach, due to the forced frequency tuning, only achieves a very low DR of approximately 0.21\% (orange). The Q-learning approach, on the other hand, can achieve a higher DR by learning the structure of the data. It searches more often in the first three sub-spectra and repeatedly observes recognized interference signals. The DR is higher with lower exploration (blue) than with higher exploration (cyan), because the receiver channels are less often distracted to other partial spectra. The approach with memory (green) has a higher standard deviation than the other algorithms with an average DR of approximately 41.79\%. 
	
	Regarding the visited frequency bands (Fig. \ref{fig:Analyses_Visits}), the Q-learning methods exhibit an imbalance in favor of the heavily occupied partial spectra, which is due to the repeated observation of the sub-spectrum containing the recognized interference signal. In comparison, it can be clearly seen that the linear frequency tuning has a balanced observation behavior between the sub-spectra (as expected). All spectra are visited exactly 20 times within an OS (orange). The Q-learning algorithm with low exploration mainly observes the first three sub-spectra (blue). With more exploration (cyan), the algorithm becomes similar to the heuristic approach. On average, the Q-learning algorithm with memory appears to be the most balanced one (green), although it again has a high standard deviation. The variability regarding the visited frequency bands observed for the Q-learning methods may be due to the random initialization employed for the Q-values.
	
	\section{Conclusion}
	
	\noindent In this paper, a significantly higher detection rate of interference signals could be achieved with a trained Q-learning algorithm compared to a heuristic approach employing linear frequency tuning. High detection rates naturally come at the expense of lower exploration -- a trade-off which can be optimally adjusted to the signal environment of interest, using free parameters available within the Q-learning algorithm. 
	
	In future work, the Q-learning approaches must be tested with other signal types, such as pulse signals and signals with low duty cycle. For a better comparison, the Q-learning algorithm with memory might also be compared with an alternative heuristic approach which allows to dwell on detected signals. Yet, it should be noted that repeated observation of already detected interference signals always negatively affects the balance of exploration. Finally, promising ReMa approaches should be integrated into existing FSM systems to verify their advantages within field tests and realistic signal environments.
	
	\section*{Acknowledgments}
	\noindent The authors would like to thank Dr.~Alwin Reinhardt (Rohde \& Schwarz) and Dr.~Qi Wang (University of the West of Scotland) for fruitful discussions.\\
	
	\section*{References}
		\begin{itemize}
		\footnotesize
		\item[1.] M.~Mostafavi, J.M.~Niya, H.~Mohammadi, and B.M.~Tazehkand, "Fast convergence resource allocation in IEEE 802.16 OFDMA systems with minimum rate guarantee," {\it China Commun.}, vol.~13, no.~12, pp.~120--131, Dec.~2016.
		\vspace{0cm}
		\item[2.] D.~Niyato and E.~Hossain, "Radio resource management games in wireless networks: an approach to bandwidth allocation and admission control for polling service in IEEE 802.16," {\it IEEE Wireless Commun.}, vol.~14, no.~1, pp.~27--35, Feb.~2007.
		\vspace{0cm}
		\item[3.] U.~Challita, L.~Dong, and W.~Saad, "Proactive resource management for LTE in unlicensed spectrum: a deep learning perspective," {\it IEEE Trans. Wireless Commun.}, vol.~17, no.~10, pp.~4674--4689, Jul.~2018.
		\vspace{0cm}
		\item[4.] A.~Selim, F.~Paisana, J.A.~Arokkiam, Y.~Zhang, L.~Doyle, and L.A.~DaSilva, "Spectrum monitoring for radar bands using deep convolutional neural networks," in {\it Proc.~2017 IEEE Global Commun.~Conf. (Globecom)}, Dec.~2017. doi:10.1109/GLOCOM.2017.8254105
	\end{itemize}
	\end{multicols}

\end{document}